\documentclass[12pt,preprint]{aastex}

\shorttitle{The first Low-Mass Black Hole?}
\shortauthors{Gelino \& Harrison}

\begin{document}

\title{GRO J0422+32: The Lowest Mass Black Hole?\altaffilmark{1}}

\author{Dawn M. Gelino \email{dgelino@ucsd.edu}}
\affil{Center for Astrophysics and Space Sciences, Mail Code 0424, 9500 
Gilman Drive, University of California at San Diego, La Jolla, CA 92093-0424}
\author{Thomas E. Harrison \email{tharriso@nmsu.edu}}
\affil{Department of Astronomy, New Mexico State University, Las Cruces, NM 
88003}

\altaffiltext{1}{This work was based on observations obtained
with the Apache Point Observatory 3.5-meter telescope, which is owned
and operated by the Astrophysical Research Consortium, and the 3-meter Shane telescope at Lick Observatory, which is a Multi-campus Research Unit of the University of California.}

\begin{abstract}
We have obtained optical and infrared photometry of the soft X-ray transient
GRO J0422+32.  From this photometry, we find a secondary star spectral 
type of M1, and an extinction of A$_V$ = 0.74$\pm$0.09.  We present 
the first observed infrared ($J$-, $H$-, and $K$-band) ellipsoidal variations, and 
model them with WD98, a recent version of the Wilson-Devinney light curve modeling code. Assuming no significant contamination of the infrared light curves, we find a lower limit to the inclination angle of 43$^{\circ}$ corresponding to an upper limit on the mass of the compact object of 4.92M$_{\odot}$. Combining the models with the observed spectral energy distribution of the system, the most likely value for the orbital    
inclination angle is $45^{\circ}\pm2^{\circ}$.  This inclination angle corresponds 
to a primary black hole mass of $3.97\pm0.95 \,M_{\odot}$.  Thus we contend that J0422+32 contains the lowest mass stellar black hole reported, and the first to have a measured mass that falls in the 3 -- 5 M$_{\odot}$ range.
\end{abstract}

\keywords{binaries: close
$-$ stars: black holes
$-$ stars: individual (J0422+32)
$-$ stars: low mass
$-$ stars: variables: other}

\section{Introduction}

Soft X-Ray Transients (SXTs) are a subset of low mass X-ray binaries that
exhibit large and abrupt X-ray and optical outbursts separated by long 
intervals of quiescence.  In most cases the compact
object is a black hole and the companion star is a low-mass K- or M-type
dwarf (see Charles 2001 for a review). During their periods of 
quiescence, SXTs are very faint at X-ray and optical wavelengths, however, 
in this state, the secondary stars can dominate the system luminosity and
 allow us to derive the parameters of the SXT binary.

J0422+32 ($\alpha_{2000}$ = 04$^{\rm h}$21$^{\rm m}$42.8$^{\rm s}$, 
$\delta_{2000}$ = 32$^{\circ}$54$\arcmin$26.6$\arcsec$) was discovered by 
the Burst and Transient Source Experiment on the Compton Gamma Ray 
Observatory on 1992 August 5 \citep{pac92}, while the $V$$\sim$13 optical 
counterpart was identified by \citet{cas93}.  Since its initial outburst, 
J0422+32 has been studied by \citet{fil95}, \citet{bee97}, \citet{har99}, 
\citet{web00}, and references therein.  These authors have found an orbital 
period of 5.092 hr, a secondary star spectral type of M2$\pm$2, a mass ratio 
of 0.116$_{-0.071}^{+0.079}$, and secondary star radial velocity 
semi-amplitude of 372$\pm$10 km s$^{-1}$.  The implied mass function for 
J0422+32 is 1.13$\pm$0.09 M$_{\odot}$ \citep{har99}, well below that of the 
accepted maximum mass of a neutron star ($\sim$ 3.2 M$_{\odot}$).

The X-ray and optical outburst light curves of J0422+32 resemble those of
 the SXT prototype, V616 Mon \citep{kin96}, however, the source has 
exhibited recurring optical and X-ray mini-outburst behavior since 1992, reminiscent of X-ray binaries with neutron star primaries (J0422+32 is not alone in this respect, however). In order to determine the nature of 
the primary object in J0422+32, its mass, and therefore orbital 
inclination angle, must be accurately measured.  As discussed in 
\citet[hereafter Paper I]{gel01} and \citet[hereafter Paper II]{gho01}, the 
best way to find the inclination angle in a non-eclipsing system is to model 
its infrared ellipsoidal light curves.  Previously, the only ellipsoidal 
variations detected from J0422+32 have been in the optical 
\citep{ca96a, che96, cea95, oro95}.  One previously published attempt was made to observe the variations in the 
infrared, where there is a smaller chance of contamination from other sources 
of light in the system, but even after combining the $H$- and $K$-band data, 
no ellipsoidal variations were seen \citep{bee97}.  Nonetheless, orbital 
inclination angle estimates have still been made.  These previously published 
inclination angles for J0422+32 range from 10$^{\circ}$\ [lower limit from 
\citet{bee97}] to 51$^{\circ}$\ [upper limit from \citet{fil95}], and 
correspond to primary masses of 34.8 M$_{\odot}$ and 2.4 M$_{\odot}$, 
respectively.  In addition, \citet{bon95} compared the intrinsic optical emission line widths of J0422+32 to those of V616 Mon, and found that the J0422+32 primary is a neutron star with M$_1 \le$ 2.2 M$_{\odot}$.

In order to determine the mass, and therefore the nature, of the compact object in J0422+32, we have obtained $J$-, $H$-, and $K$-band light curves and model them here with WD98, a recent version of the Wilson-Devinney light curve modeling code.  In Section 2 we describe our observations and data reduction process, as well as present the infrared photometric light curves.  We describe our choices for the relevant WD98 input parameters and present our models of the variations in Section 3. Finally, in Section 4, based on the presented models, we discuss the nature of the compact object in J0422+32. 

\section{Observations \& Data Reduction}

J0422+32 was observed using GRIM II\footnote{See 
http://www.apo.nmsu.edu/Instruments/GRIM2/} on the Astrophysical Research 
Consortium 3.5 meter telescope at Apache Point Observatory on 2000 January 23, 
24, 25, and 26.  $J$-band data were obtained on all of these nights, while $K_s$-band data were obtained only on 2000 January 25 and 26.  The data were linearized before averaging the images at one 
position and subtracting them from the average of the images at the other 
position.  The images were flat fielded with a dome flat using the usual 
IRAF\footnote{IRAF is distributed by the National Optical Astronomy 
Observatories, which are operated by the Association of Universities for 
Research in Astronomy, Inc., under cooperative agreement with the National 
Science Foundation.} packages. 

The SXT was observed again on 2003 February 7 and 8 using the Gemini 
Twin-Arrays Infrared Camera\footnote{See http://mthamilton.ucolick.org/techdocs/instruments/gemini/gemini$\_$index.html} on the Shane 3 meter 
telescope at Lick Observatory. Simultaneous $J$- and $K'$-, and $H$- and 
$K'$-data were obtained.  The number of counts in each exposure was kept 
in the linear regime of each chip, as there is no linearization correction 
available.  These data were reduced in the same manner as above using 
twilight flats.

\begin{figure}
\plotone{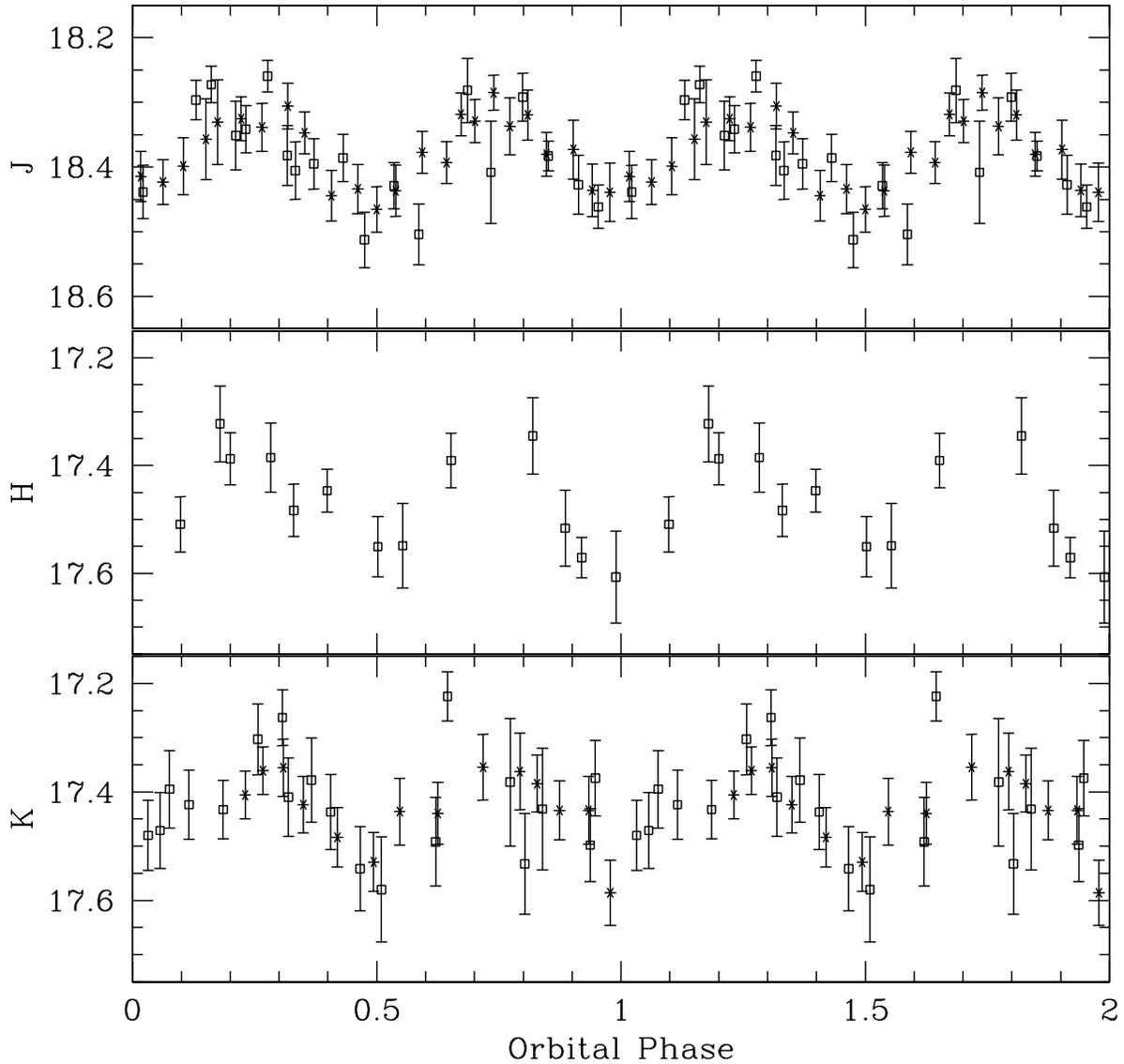}
\caption{Top panel: Long-term infrared light curves of J0422+32.  The $J$- and $K$-band data span 37 months.  The asterisks represent data taken January 2000  with the Astrophysical Research Consortium 3.5 m telescope, and the open squares represent data taken February 2003 with the Lick Observatory Shane 3 m telescope.  Error bars are 1-$\sigma$. These data are plotted over two phase cycles for clarity. Here, and throughout this paper, we phase the J0422+32 heliocentric-corrected data to the ephemeris of \citet{web00}.  There are no obvious long-term variations in the shapes or mean magnitudes of the light curves. 
\label{fig1}
}
\end{figure}

Aperture photometry was performed on J0422+32 and five nearby field stars.
  Using the IRAF PHOT package, a differential light curve for each band was generated seperately for the 2000 and 2003 data sets, 
with each point being the average of four beam switched images. The 
differential photometric results show that over the course of our 
observations, the comparison stars did not vary more than expected from 
photon statistics.  Variations were observed in both
the 2000 and 2003 data sets.  Figure \ref{fig1} compares the 2000 and 2003 light curves.

\begin{figure}
\plotone{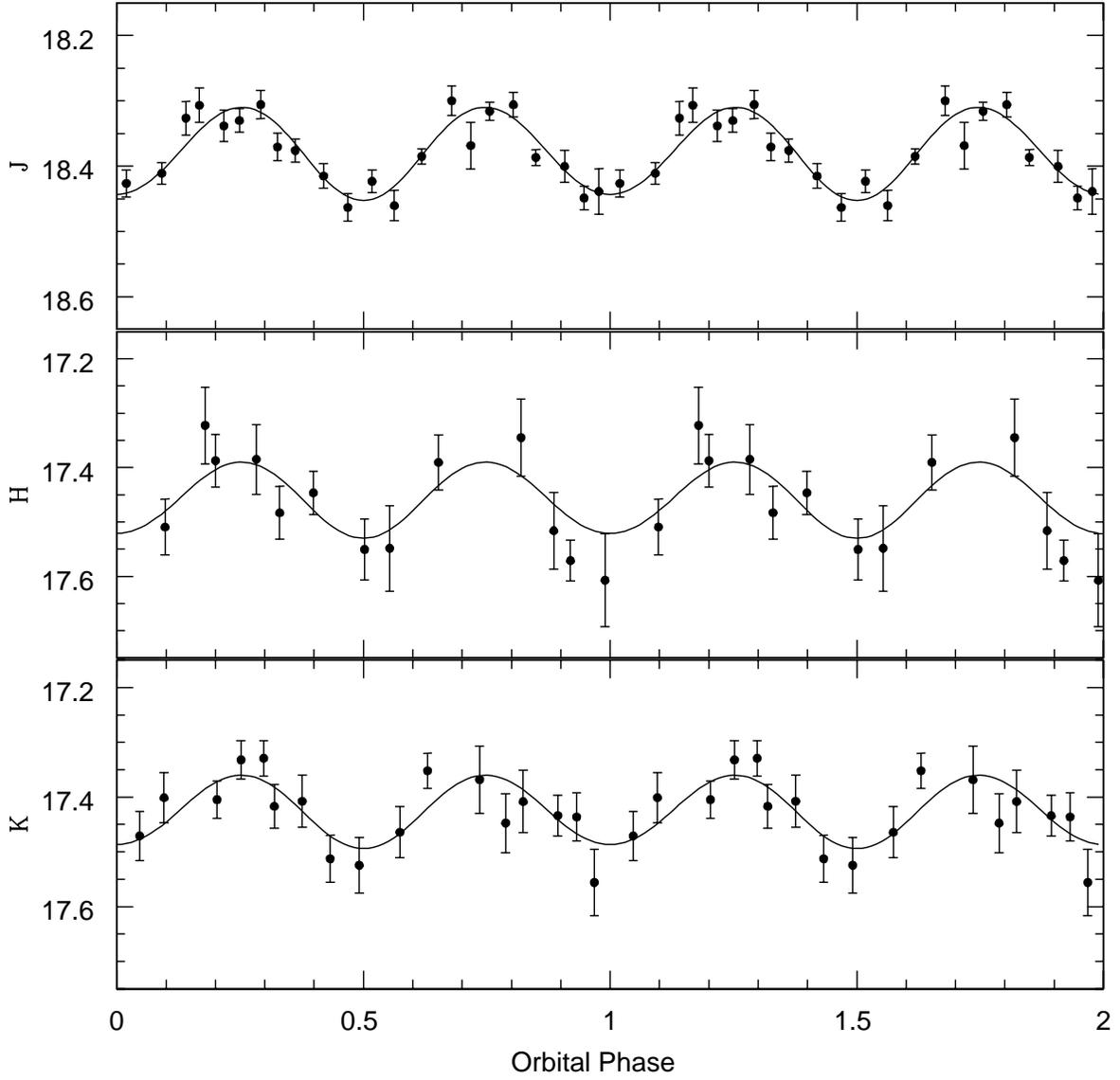}
\caption{J0422+32 $J$-, $H$-, and $K$-band final light curves (points).  Error bars are 1-$\sigma$. The solid line represents the best fitting ($i$ = 45$^{\circ}$) WD98 model as described in the text. 
\label{fig2}
}
\end{figure}

As Figure \ref{fig1} shows, the 2000 and 2003 observations revealed light curve shapes, amplitudes, and brightnesses which were consistent with each other. This enabled us to combine both data sets.  The final $J$, $H$, and $K_s$ differential 
light curves of J0422+32, phased to the ephemeris of \citet{web00}, are 
presented in Figure \ref{fig2}. These light curves represent the first 
detections of infrared ellipsoidal variations from this SXT.

Optical observations of J0422+32 were
 obtained with SPIcam\footnote{See 
http://www.apo.nmsu.edu/Instruments/SPIcam/} on 2001 December 18.  The 
$B$-, $V$-, $R$-, and $I$-band exposures were bias-corrected and flat-fielded 
before aperture photometry was performed. Standards were also observed to transform these data to the system of Landolt. The colors of J0422+32 can be 
found in Table~\ref{tab1}, along with other quiescent optical and infrared
 colors published for this SXT.

\section{Infrared Light Curve Modeling}

The J0422+32 infrared light curves presented in Figure \ref{fig2} were 
modeled with WD98 in order to find the orbital inclination angle of the 
system.  Many of the parameters needed for light curve modeling are based 
upon the nature of the secondary star. Before we can model 
the infrared light curves of J0422+32, we must first derive the input
model parameters, and we do so here. 

\subsection{The Nature of the Secondary Star in J0422+32}

The most important parameter for modeling the infrared light curves of
any non-eclipsing system is the spectral type of the secondary star.
This property can be estimated by comparing red optical spectra of the SXT secondary star with various spectral types from the same luminosity class. Because
of the orbital motion of J0422+32 ($\pm$ 372 km s$^{-1}$), the photospheric absorption
lines are broadened by rotation. This effect is usually accounted for during the analysis of the spectral data. The optical spectrum may also suffer from
contamination. Both effects can weaken the apparent line strengths,
and make stellar classification difficult. Another difficulty is that to
classify the secondary star in these systems, it is assumed that the gravity
and abundances of the secondary star are consistent with those of the main 
sequence templates used to derive the spectral type. Given the low S/N of these
data, deriving a spectral type can be challenging.

\begin{figure}
\plotone{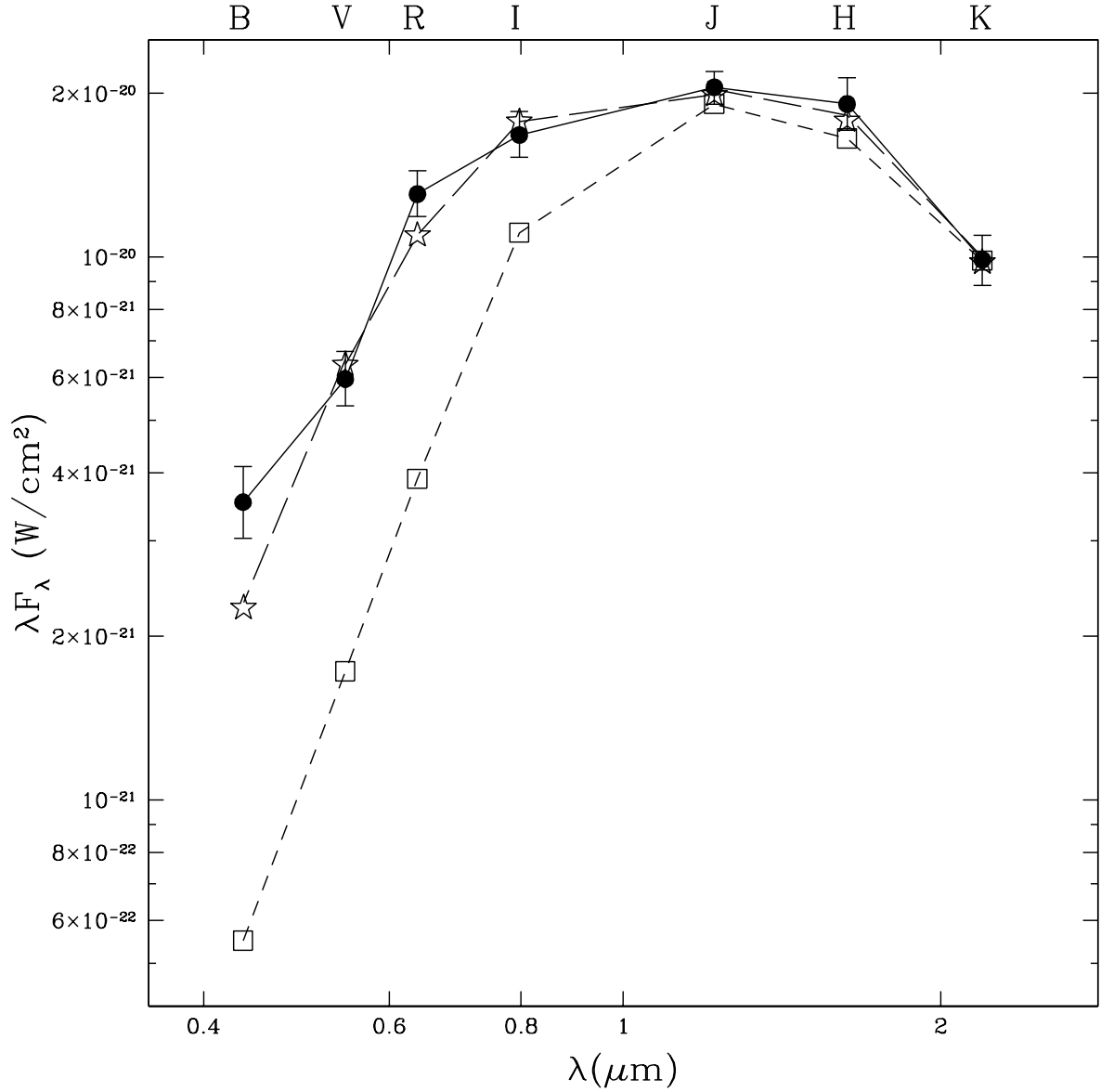}
\caption{J0422+32 optical/infrared quiescent SED dereddened by A$_{\rm V}$ = 0.74 mag (filled circles).  The infrared ($JHK$) points represent phase-averaged data.  Error bars are $1\sigma$. These data are compared to the SED for an M1V (open stars), and an M4V (open squares), normalized at $K$.  The M1V SED fits that of J0422+32 with exceptions at $B$ (34\%) and $R$ (8\%).
\label{fig5}
}
\end{figure}

Alternatively, one can use a spectral energy distribution (SED), and the above
limits on the spectral type to not only derive an effective temperature of
the secondary star, but also estimates for both the visual extinction and 
contamination level. Although difficulties similar to finding the spectral type of the secondary star from spectra may be encountered when using broad-band photometry, the effective temperature is less sensitive to changes in the gravity and abundances than the shape and depth of individual spectral features.
Given the higher S/N an effective temperature derived
using photometry can be just as useful as a spectral type derived from the 
spectroscopic data set. We present the optical-infrared ($BVRIJHK$) SED for 
J0422+32 in Figure \ref{fig5}. In this figure we compare the observed SED to those for an 
M1V and an M4V \citep{bb88,bes91,mik82}.  We find that after dereddening the observed data by A$_V$=0.74
mags, the best-fitting spectral type is that of an M1V. This extinction is
consistent with published values. \citet{shr94} calculate E(B--V)=0.2$\pm$0.1 from the 5780\AA~diffuse interstellar feature, E(B--V)=0.23$\pm$0.02 from an optical continuum power law fit, and finally E(B--V)=0.40$\pm$0.06 from an ultraviolet 2170\AA~interstellar feature.  Our color excess of E(B--V)=0.24$\pm$0.03 from SED fitting is consistent with the optically derived excess from \citet{shr94}, as well as those from \citet{kin96}, \citet{cal95}, and \citet{mar95}.   Based on the results
 from Figure \ref{fig5}, we adopt an extinction value of A$_V$=0.74$\pm$0.09 mag, a secondary star spectral type of M1, and corresponding temperature of $T_{\rm eff}$ = 3900 K \citep{gra92}.

An M1V spectral type is consistent with those found by \citet{cea95} and \citet{har99}, but not with
the M4V type arrived at by \citet{web00}. A spectral type of M4V cannot be made
consistent with the observations, no matter the extinction. The only way to 
make a spectral type as late as M4 work with the observations is to posit
the existence of an additional source of blue luminosity in the system. However, as 
is clear from the SED, such a source would have to supply 70\% of the flux in 
the $R$-band, and would dominate the red optical spectra. We address the
issue of contamination in the next section.

It can be seen from Table 1 that the infrared data presented by \citet{bee97} are
not consistent with our photometry. While the optical magnitudes and colors
have shown very little evidence for change since J0422+32 entered
quiescence, the infrared luminosity of the source appears to have declined
by more than a magnitude! By combining data from \citet{cea95} with that of
\citet{bee97} taken within the same month, we find that $V-K$ = 6.12 in 1994. 
If we assume that A$_V$=0.74, the dereddened value is ($V-K$)$_{\circ}$ = 5.46. This corresponds to a spectral type of M4/5V. Using the dereddened color from \citet{cea95}, ($V-I$)$_{\circ}$ = 1.72, a spectral type of M0V $\pm$ 1 is derived. These two colors are completely inconsistent: the $V-I$ color of an M4/5V would is 3.0 $\pm$ 0.2! This should be compared with our values of ($V-I$)$_{\circ}$ = 1.96
(M1V), and ($V-K$)$_{\circ}$ = 3.95 (M1V). 

It seems difficult to understand this type of change. We have examined the 
2MASS values for the secondary standards used by \citet{bee97}, and have
been unable to reproduce their photometry. First, in the notes
of their Table 2, they appear to have confused 
the identification of their two standard stars with those of \citet{che95}. \citet{che95} star \#4 is the brighter of the two stars, and must correspond
to their ``Standard-1'' (although it is possible that star \#4 is variable, no evidence to this effect is found in our long-term data). Even with the correct identification, however, their
photometry does not agree with the 2MASS values. If there was simply a zero point offset, the differences between the two sets of magnitudes would be consistent.  We do not find any such consistency. We are unable to determine the source of this problem, but there does appear to be something unusual about
the \citet{bee97} infrared photometry that extends beyond a mere zero point 
correction.

We conclude that the spectral type of the secondary star is M1V+/-1, consistent
with all of the previous assignments, except that of \citet{web00}. While the
methodology used by \citet{web00} is robust for normal, late-type main sequence 
stars, it is unclear whether the secondary stars in SXTs can be considered
to be normal main sequence stars. Again, broad-band photometry is not as sensitive to changes in gravity and abundances as spectral features, however we do use main sequence data in the SED and atmosphere models (see section 3.3). Therefore, for completeness, we have run models
where the spectral type of the secondary star was assigned an effective 
temperature of an M4V, the spectral type determined by \citet{web00}.  We find that our result does not significantly change with this variation in spectral type.

\subsection{Estimating the Amount of Infrared Non-Stellar Light}

If we assume a spectral type of M1V and A$_V$=0.74 mags, there is very
little room in the $VRIJHK$ SED for contamination from other sources in the
system. Because the level of contamination is an extremely important quantity,
acting to dilute the elliposidal variations, we address the issue here. There
have been a number of esitmates for the level of the $R$-band contamination in 
the J0422+32 system ranging from less than 20\% \citep{ca96a} to up
to 60\% \citep{cea95, fil95}. In our SED, we find that there
is about an 8\% {\it excess} in the $R$-band that could be attributed to H$\alpha$ emission. Estimating the level of contmaination in an optical spectrum
that has limited wavelength coverage is more difficult than spectral type
determination--it suffers from many of the same issues, while attempting to
fit a continuum to the observed spectrum.

It is clear from the SED that combinations of a hot and a cool source can
reproduce the observed spectrum.  But note that whatever
``hot'' source (anything hotter than an M1V) one selects so as 
to reproduce the $BVRI$ data by combining it with a very late-type dwarf, it 
will have only minor effects in the infrared. No one has proposed 
a spectral type later than M4 for the secondary star. If one chooses an M4V 
spectral type for the companion, and normalize the SED to the observed 
$K$-band flux, then the secondary star only supplies 30\% of the observed $R$
-band flux. 
This fraction, of course, is even smaller if you don't normalize to the $K$-band 
flux. Thus, the level of contmaination in the $R$-band using an M4 secondary
star, 70\%, is larger than any of the existing estimates for this contamination. The 
optical spectrum of such a heavily contaminated source would be highly diluted, 
and quite difficult to extract. As one adjusts the spectral type of the 
companion to earlier types, the dilution is lessened. An M2V would have a 
dilution of 24\% at $R$, with the star contributing 76\% of the flux.

The most difficult contamination source to extract in the case of J0422+32
is the one that has the {\it shallowest spectral slope}. The most relevant
emission process with the shallowest slope is free-free emission, which in 
$\lambda$F$_\lambda$ space has a spectrum proportional to $\lambda^{-1}$. If we ascribe the
entire B-band excess (34\%) to free-free emission, then in the $K$-band, the
contamination would be $\sim$ 7\%. Such a low-level of contamination would be
difficult to detect given the S/N of the current data. {\it Any other proposed process, such as synchrotron emission, or classical accretion disk spectra produces a steeper spectrum in $\lambda$F$_\lambda$ than free-free emission, and thus will dilute the infrared ellipsoidal variations by a smaller amount than free-free emission.}

A very conservative estimate of the disk contamination in the infrared is based on the assumption that the optically thin disk indeed radiates through free-free emission processes. A 7\% contamination in the infrared bands would cause the 
observed orbital inclination angle of the system to be underestimated by 
$\le$ 2$^{\circ}$.

\subsection{Ellipsoidal Models}

As in Papers I and II, we modeled the infrared light curves of J0422+32 
with WD98 \citep{wil98}.  See Paper I for references and a basic 
description of the code, and \citet{dmg01} for a more comprehensive 
description. We ran WD98 for a semi-detached binary with the primary having 
such a large gravitational potential, that it essentially has zero radius.  
The most important adopted wavelength-independent input values to WD98 are 
listed in Table~\ref{tab2}, and the wavelength-dependent values are listed in Table~\ref{tab3}.

The models were run for a range of inclination angles with parameters 
for an M0V through an M4V secondary star.  The secondary star atmosphere was 
determined from solar-metallicity Kurucz models.  The Kurucz atmosphere models
 are computed in temperature steps of 250~K for the range we are interested in, and stop at 3500~K.  We therefore use extrapolated fluxes for the cooler parts of the star.  For reasons cited above, we have used normal, non-irradiated, limb darkening coefficients in the models 
\citep{van93}.  We also adopted gravity darkening exponents found by 
\citet{cla00}, and a mass ratio, $q$, of 0.116$^{+0.079}_{-0.071}$ 
\citep{har99}.  We ran models assuming the secondary star was the only source 
of {\it infrared} light in the system.  We also assumed that the secondary 
star had a uniform surface brightness aside from limb- and gravity-darkening 
effects (i.e. no star spots). 

    The best fit model was 
determined using $\chi^2$ tests.  We found that changing the spectral type
 of the secondary from an M4V to an M0V, resulted in a change in the orbital 
inclination angle of $<$1$^{\circ}$. Similarly, varying $q$ from 0.045 to 0.195 affected $i$ by $\le 1^{\circ}$.  We find that the best fitting $J$-, $H$-, and 
$K_s$-band model has $i$ = 45$^{\circ}$, and the parameters found in Table \ref{tab2}.  Figure \ref{fig2} presents this best 
fitting model for the $J$-, $H$-, and $K_s$-band light curves.

\subsection{The Adopted Model and its Uncertainties}

Assuming all of the infrared light in the system comes from the secondary
 star, the best fitting orbital inclination angle for J0422+32 is 
$i$ = 45$^{\circ}$.  If this assumption was incorrect, and significant infrared contamination were present in the system, each light curve would most likely be diluted by a different amount, causing the best fit inclination for each band to be different from the others.  We instead find that all three infrared light curves give a consistent 45$^{\circ}$ best fit whether solved simultaneously or individually.  

If there is no contamination affecting the infrared light curves, the absolute lower limit to the inclination angle is 43$^{\circ}$.  This corresponds to an upper limit to the compact object mass of 4.92 M$_{\odot}$, firmly placing it below 5 M$_{\odot}$.  Using an estimate of 7\% for the infrared accretion disk 
contamination in the system gives an inclination of 
45$^{+2.8}_{-2}$$^{\circ}$, however, we do not believe that the infrared 
light curves are significantly affected by any such contamination.  
Therefore, based on the error in $q$, the spectral type (i.e. 
temperature) of the secondary star, and the photometric error bars, the 
adopted orbital inclination angle is 45$^{\circ}$$\pm$2$^{\circ}$.  
Combining this inclination with the observed mass function, we find the mass 
of the primary object to be 3.97$\pm$0.95 M$_{\odot}$. 

Using the mass of the compact object and the orbital period, the orbital separation of the two components in the system can be computed.  This in turn, can be combined with the mass ratio to find the size of the Roche lobe for the secondary star.  The temperature of the secondary and its Roche lobe radius are then used to find the secondary's bolometric luminosity and bolometric absolute magnitude. After accounting for the bolometric correction \citep{bes91}, the distance modulus for the $J$, $H$, and $K$ bands is used to find an average distance of 2.49$\pm$0.30 kpc. Table~\ref{tab4} lists all of the derived parameters for J0422+32.

\section{Discussion}

In this paper, we have presented the first observed infrared ellipsoidal 
variations for J0422+32.  The derived parameters in Table~\ref{tab4} are 
based on the modeling of these variations assuming no significant 
contamination from any other sources of infrared light in the system.  Even though there is a large range of published inclination angles for this system, the inclination angle found here is consistent with those found by \citet[$\ge$ 45$^{\circ}$]{oro95}, \citet[45$^{\circ}$--51$^{\circ}$]{fil95}, \citet[$<$ 45$^{\circ}$]{ca96a}, and \citet[$\le$ 45$^{\circ}$]{web00}, and suggests that we should not be overly surprised that this system harbors a low-mass object.  

Even though the errors determined here for the orbital inclination angle and mass of the primary object are smaller than previously published, the nature of the 
primary object in this system remains unclear.   Both the optical and X-ray 
outburst light curves of J0422+32 closely resemble those of the prototype
 black hole SXT, V616 Mon.  However, since outburst, the system has exhibited 
a series of mini-outbursts, similar to those observed from neutron star 
X-ray binaries such as Aql X-1 \citep{cha80}.  \citet{mar95} not only compare 
this SXT with Aql X-1, but also compare J0422+32 to GX 339-4, an X-ray binary 
system thought to be a black hole candidate based on its rapid X-ray 
variability, and because it exhibits ``high,'' ``low,'' and ``off'' states 
\citep{mar73}.  They conclude their study by saying, ``J0422+32 would seem 
to be a system whose nature is intermediate between the X-ray transients and 
those which show more than one mode of accretion.''

 We find the minimum mass of the primary object to be just below the 
maximum accepted mass for a neutron star.  However, just about all neutron 
stars with measured masses cluster around 1.35 M$_{\odot}$ with a small 
spread of $\pm$0.04 M$_{\odot}$ \citep{van01}.  For the compact object in this system to have a mass consistent with that of a typical neutron star, the orbital inclination angle would have to be about 85$^{\circ}$, which is ruled out by a lack of eclipses during outburst. In addition, contamination at the 160\% level in the infrared is needed to account for a 40$^{\circ}$ underestimation of the inclination angle.  This level of contamination surely would be visible in the SED, and rule out any possibility of detecting
the secondary star in the optical, therefore until more compelling evidence 
for a neutron star primary is presented, we suggest that the primary object
 in J0422+32 is a low mass black hole.  

Recently, two papers (Paper II; \citet{gre01}) have published evidence 
for the highest-mass stellar black holes. Here we find what appears to be
 the lowest mass stellar black hole.  A handful of other sources may be
 potential candidates based on their observed mass functions, but current 
findings suggest otherwise.  The most likely candidate to join J0422+32 as 
a low mass black hole is GRS 1009-45. The current estimates of the compact object mass span from 4.64M$_{\odot}$ to 5.84 M$_{\odot}$ \citep{dmg02} with a lower limit of 4.4M$_{\odot}$ from lack of X-ray eclipses \citep{fil99}.  Thus we contend that 
J0422+32 contains the lowest mass stellar black hole, and the first to 
have a measured mass that falls in the 3 -- 5 M$_{\odot}$ range.

\acknowledgments

We would like to thank John Tomsick for helpful discussions, and the Apache Point Observatory and Lick Observatory staff, especially Elinor Gates.  DMG held an
 American Fellowship from the American Association of University Women 
Educational Foundation while working on this project in 2000, and currently holds a CASS Postdoctoral Fellowship.

\clearpage
\begin{deluxetable}{cccccccc}
\tabletypesize{\scriptsize}
\tablewidth{0pt}
\tablenum{1}
\tablecaption{Quiescent Infrared and Optical Colors of J0422+32\label{tab1}}
\tablehead{
\colhead{Reference} &\colhead{V} &\colhead{V - R}
&\colhead{V - I} &\colhead{K} &\colhead{J - K} &\colhead{H - K} &\colhead
{Date\tablenotemark{a}}
}
\startdata
1 & 22.39$\pm$0.27 & 1.33$\pm$0.29 & ... & ... & ... & ... & 9/93 \\
2 & 22.24$\pm$0.14 & 1.27$\pm$0.17 & 2.02$\pm$0.16 & ... & ... & ... & 12/94 \\
3 & ... & ... & ...  & 16.12$\pm$0.04\tablenotemark{b} & ... & 0.31$\pm$0.05\tablenotemark{b} & 12/94 \\
4 & 22.3$\pm$0.2 & 1.3$\pm$0.2 & ... & ... & ... & ... & 9/94--1/95 \\
5 & 22.35$\pm$0.17 & 1.41$\pm$0.20 & ... & ... & ... & ... & 9/94--3/95 \\
6 & 22.05$\pm$0.10\tablenotemark{c} & 1.31$\pm$0.12 & 2.26$\pm$0.12 & 17.44$\pm$0.09 & 0.95$\pm$0.10 & 0.16$\pm$0.16 & 2000/2001/2003 \\
&&&&&&& \\
\enddata
\tablenotetext{a}{Month and year the data were obtained}
\tablenotetext{b}{Data cannot be reconciled with 2MASS, see text}
\tablenotetext{c}{B-V=1.17$\pm$0.17}
\tablerefs{(1) \citet{zha94} (2) \citet{cea95} (3) \citet{bee97} (4) \citet{ca96a} (5) \citet{gar96} (6) This Paper}
\end{deluxetable}{}
          
\newpage    
\begin{deluxetable}{lc}
\tablenum{2}
\tablecaption{Wavelength-Independent WD98 Input Parameters\label{tab2}}
\tablehead{
\colhead{Parameter} &\colhead{Value}
}
\startdata
Orbital Period (days) & 0.2121600 \\
Ephemeris (HJD phase 0.0)\tablenotemark{a} & 2450274.4156\\
Orbital Eccentricity & 0.0 \\
Temperature of Secondary (K) & 3900 \\                 
Mass Ratio (M$_2$/M$_1$) & 0.116 \\
Atmosphere Model & Kurucz (log g = 4.59)\\
Limb Darkening Law & Square-root \\ 
Secondary Star Gravity Darkening Exponent & $\beta_1$=0.27 \\
Secondary Star Bolometric Albedo & 0.676 \\
\enddata
\tablenotetext{a} {From \citet{web00}}
\end{deluxetable}{}

\newpage    
\begin{deluxetable}{lccc}
\tablenum{3}
\tablecaption{Wavelength-Dependent WD98 Input Parameters\label{tab3}}
\tablehead{
\colhead{Parameter} &\colhead{J}&\colhead{H}&\colhead{K}
}
\startdata
Square-root Limb Darkening Coefficient x$_\lambda$& -0.137 & -0.146 & -0.160 \\
Square-root Limb Darkening Coefficient y$_\lambda$& 0.718 & 0.655 & 0.592 \\
\enddata
\end{deluxetable}{}

\newpage
\begin{deluxetable}{lc}
\tablenum{4}
\tablecaption{Derived Parameters for J0422+32\label{tab4}}
\tablehead{
\colhead{Parameter}&\colhead{Value}}
\startdata
Orbital Inclination Angle ($^{\circ}$) & 45$\pm$2 \\
Primary Object Mass $M_1$ ($M_{\odot}$) & $3.97\pm 0.95$ \\
Secondary Star Mass $M_2$ ($M_{\odot}$) & $0.46\pm 0.31$ \\
Orbital Separation $a$ ($R_{\odot}$) & $2.45\pm 0.24$ \\
Secondary Star Radius $R_{L_2}$ ($R_{\odot}$) & $0.53\pm 0.16$  \\
Adopted Distance (kpc) & $2.49\pm 0.30$ \\
\enddata
\end{deluxetable}{}

\clearpage

\end{document}